\documentclass[aip,cha,reprint]{revtex4-1}

\usepackage{amssymb}
\usepackage[english]{babel}   
\usepackage[dvips]{graphics}
\usepackage{subfigure}
\usepackage[pdftex]{graphicx}
\usepackage{epsfig}
\usepackage{multirow}
\usepackage{rotating}
\usepackage{picinpar}
\usepackage{color}

\begin{document}

\title{TEXTURE ANALYSIS AND CHARACTERIZATION USING PROBABILITY FRACTAL DESCRIPTORS}
\author{Jo\~{a}o B. Florindo}
\email{jbflorindo@gmail.com}
\affiliation{Instituto de F\'{i}sica de S\~{a}o Carlos (IFSC) Universidade de S\~{a}o Paulo, Av.  Trabalhador S\~{a}o Carlense, 400,
CEP 13560-970 - S\~{a}o Carlos, S\~{a}o Paulo, Brasil,phone/fax: +55 16 3373 8728 / +55 16 3373 9879} 

\author{Odemir M. Bruno}
\email{bruno@ifsc.usp.br}
\affiliation{Instituto de F\'{i}sica de S\~{a}o Carlos (IFSC) Universidade de S\~{a}o Paulo, Av.  Trabalhador S\~{a}o Carlense, 400,
CEP 13560-970 - S\~{a}o Carlos, S\~{a}o Paulo, Brasil,phone/fax: +55 16 3373 8728 / +55 16 3373 9879}       

\date{\today}

\begin{abstract}
A gray-level image texture descriptors based on fractal dimension estimation is proposed in this work. The proposed method estimates the fractal dimension using probability (Voss) method. The descriptors are computed applying a multiscale transform to the fractal dimension curves of the texture image. The proposed texture descriptor method is evaluated in a classification task of well known benchmark texture datasets. The results show the great performance of the proposed method as a tool for texture images analysis and characterization.
\end{abstract}


\maketitle



\section{Introduction}

Fractals have played an important role in many areas with applications related to computer vision and pattern recognition \cite{SMS10,HWZ08,CDHLAB03,W08,TWZ07,LC10,RossattoCKB11,FlorindoSPB2012}. This broad use of fractal geometry is explained by the flexibility of fractals in representing structures usually found in the nature. In such objects, we observe a level of details at different scales which are described in a straightforward manner by fractals, rather than through classical Euclidean geometry.

Most of fractal-based techniques are derived from fractal dimension concept. Although this concept was defined originally only for real real fractal objects, it contains some properties which turn it into a very interesting descriptor for any object of real world. Indeed, the fractal dimension measures how the complexity (level of details) of an object varies along scales. Such definition corresponds to an effective and flexible means of quantifying the spatial occupation of an object as well as physical and visual important aspects which characterize an object, like luminance and roughness, for instance.

Among these fractal techniques, we may cite Multifractals \cite{H01,LRAJ08,LGS00}, Multiscale Fractal Dimension \cite{MCSM02,CC00,PPFVOB05} and Fractal Descriptors \cite{BPFC08,BCB09,FCB10,Florindo2012798,Backes20121984,FlorindoB11}. Here we are focused on the later approach which demonstrated the best results in such kind of application \cite{FlorindoB12}. The main idea of fractal descriptors theory is to provide descriptors of an object represented in a digital image from the relation between the values of fractal dimension taken under different observation scales. These values provide a valuable information about the complexity of the object in the sense that they capture the degree of details at each scale. In this way, fractal descriptors are capable of quantifying important physical characteristics of the structure, like fractal dimension, but presenting a sensibly richer information than a unique number (fractal dimension).

Here, we proposed a novel fractal descriptors based on probability fractal dimension. We used the whole power-law curve of the dimension and applied a space-scale transform to emphasize the multiscale aspect of the features. Finally, we test the proposed method over two well-known datasets, that is, Brodatz and Vistex, comparing the results with another fractal descriptors approach showed in \cite{BCB09} and other conventional texture analysis methods. The results demonstrated the power of probability descriptors, achieving a more precise classification than other classical techniques.

\section{Fractal Theory}
 
In the last decades we observe a growing number of works applying fractal geometry concepts in the solution of a wide range of problems \cite{SMS10,HWZ08,CDHLAB03,W08,TWZ07,LC10}. This interest in fractals is mainly motivated by the fact that conventional Euclidean geometry has severe limitations in providing accurate measures of real world objects.

\subsection{Fractal Dimension}

The first definition of fractal dimension provided in \cite{M68} is the Hausdorff dimension. In this definition, we consider a fractal object as being a set of points immersed in a topological space. Thus, we can use results from Measure Theory to define a measure over this object. This is the Hausdorff measure expressed through:
\[
	H^{s}_{\delta}(X) = \inf{\sum_{i=1}^{\infty}{|U_{i}|^{s}\mbox{: {U$_{i}$ is a $\delta$-cover of X}}}},
\]
where $|X|$ denotes the diameter of $X$, that is, the maximum possible distance between any elements of $X$:
\[
	|X| = \sup\{|x-y|:x,y \in X\}.
\]
Here, we say that a countable collection of sets ${U_i}$, with $|U_i| \leq \delta$, is a $\delta$-cover of $X$ if $X \subset \cup_{i=1}^{\infty}U_i$.

Notice that $H$ also depends on a parameter $\delta$ which expresses the scale under which the measure is taken. We can eliminate such dependence by applying a limit over $\delta$, defining in this way the $s$-dimensional Hausdorff measure:
\[
	H^s(X) = \lim_{\delta \rightarrow 0}H^s_{\delta}(X).
\]
If we plot $H^s(X)$ as a function of $s$ we observe a similar behavior in any fractal object analyzed. The value of $H$ is $\infty$ for any $s < D$ and it is $0$ for any $s > D$, where $D$ is always a non-negative real value. $D$ is the Hausdorff fractal dimension of $X$. In a more formal way we may write:
\[
	D(X) = \{s\} | \inf \left\{ s:H^{s}(X)=0 \right\} = \sup \left\{ H^{s}(X)=\infty \right\}.
\]

In most practical situations, Hausdorff dimension uses to be complicate or even impossible to calculate. Thus, assuming that any fractal object is intrinsically self-similar, we may derive a simplified version, also called similarity dimension or capacity dimension:
\[\label{eq:FD}
	D = -\frac{\log(N)}{\log(r)},
\] 
where $N$ is number of rules with linear length $r$ used to cover the object. 

In practice, the above expression may be generalized by considering $N$ as any kind of self-similarity measure and $r$ as any scale parameter. This generalization gives rise to a lot of estimation methods for fractal dimension, with a broad application to the analysis of objects which are not real fractals (mathematically defined) but which present some degree of self-similarity in specific intervals. An example of such method is the probability dimension, used in this work and described in the following section.

\subsection{Probability Dimension}

The probability dimension, also known as information dimension is derived from the information function. This function can be defined in any situation where we have an object populating a physical space. We must divide this space into a grid of squares with side-length $\delta$ and calculate the probability $p_m$ of $m$ points of the object pertaining to some square of the grid. The probability function is given by:
\[
	N_P(\delta) = \sum_{m=1}^{N}{\frac{1}{m}p_m(\delta)},
\]
where $N$ is maximum possible number of points of the object inside a unique square.

The dimension itself is given through:
\[\label{eq:prob}
	D = -\lim_{\delta \rightarrow 0}\frac{\ln N_P}{\ln \delta}.
\]

When this dimension is estimated over a gray-level digital image $I:[M,N] \rightarrow \Re$, an usual approach is to map it onto a three-dimensional surface $S$ through:
\begin{equation}
	S = \{i,j,I(i,j)|(i,j) \in [1:M] \times [1:N]\}.
\end{equation}
In this case, we construct a three-dimensional grid of 3D cubes also with side-length $\delta$. The probability $p_m$ is therefore calculated as being the number of grid cubes containing $m$ points of the surface divided by the maximum number of points inside a grid cube.
\begin{figure*}[!htbp]
	\centering
	\begin{tabular}{cccc}
		\includegraphics[width=0.3\textwidth]{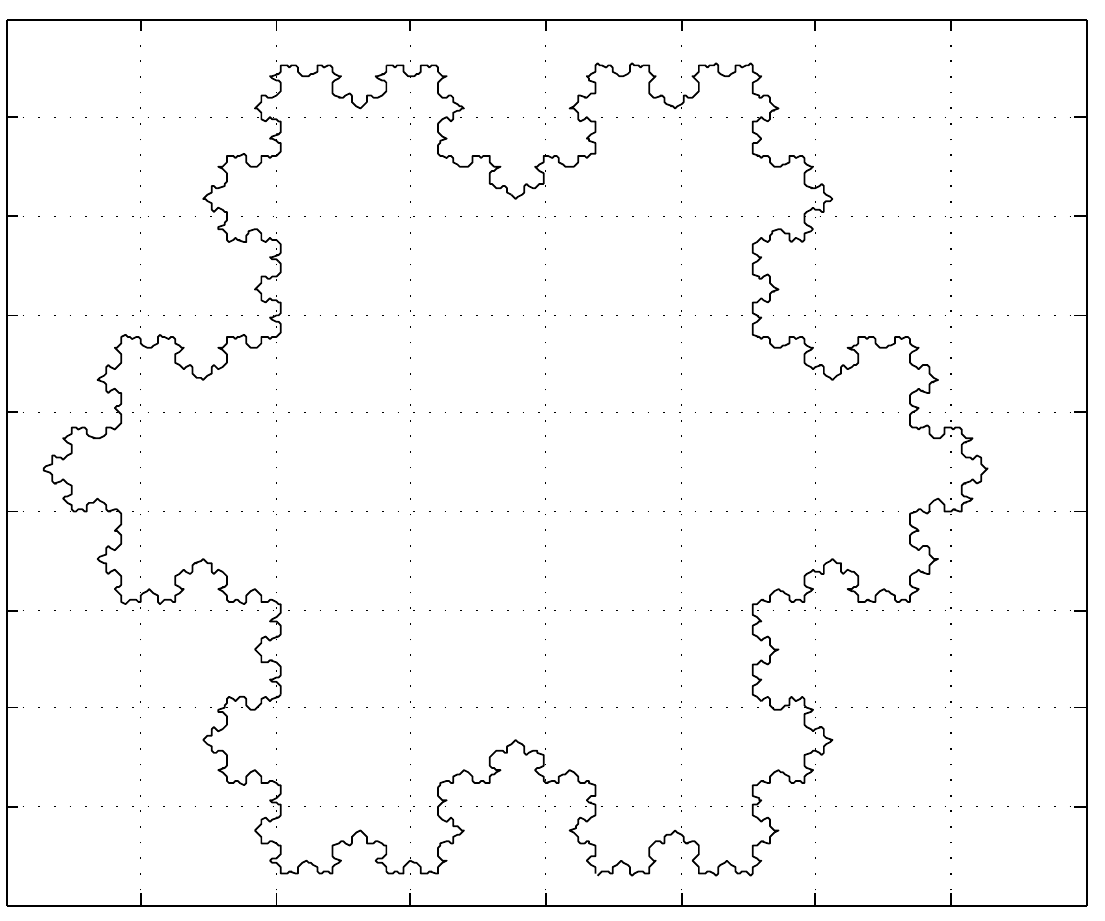} & \includegraphics[width=0.3\textwidth]{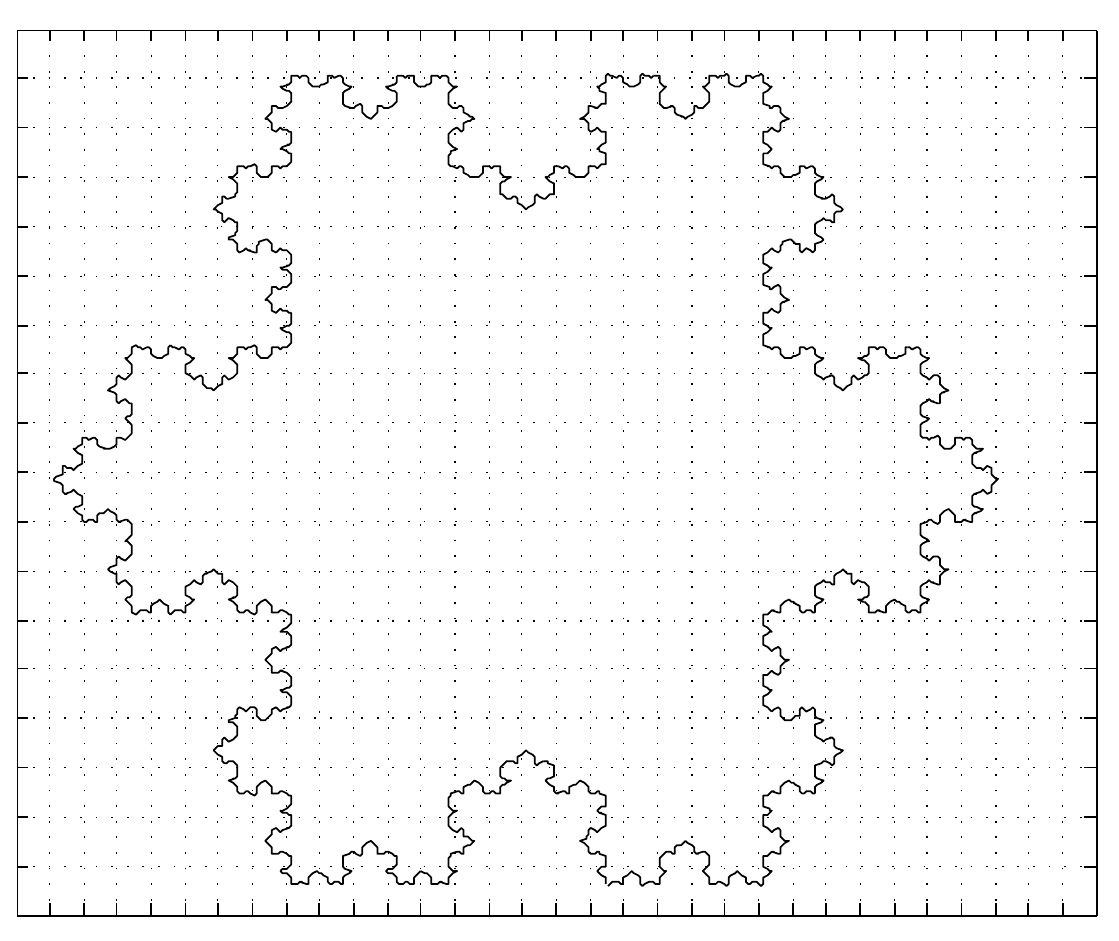} & \includegraphics[width=0.3\textwidth]{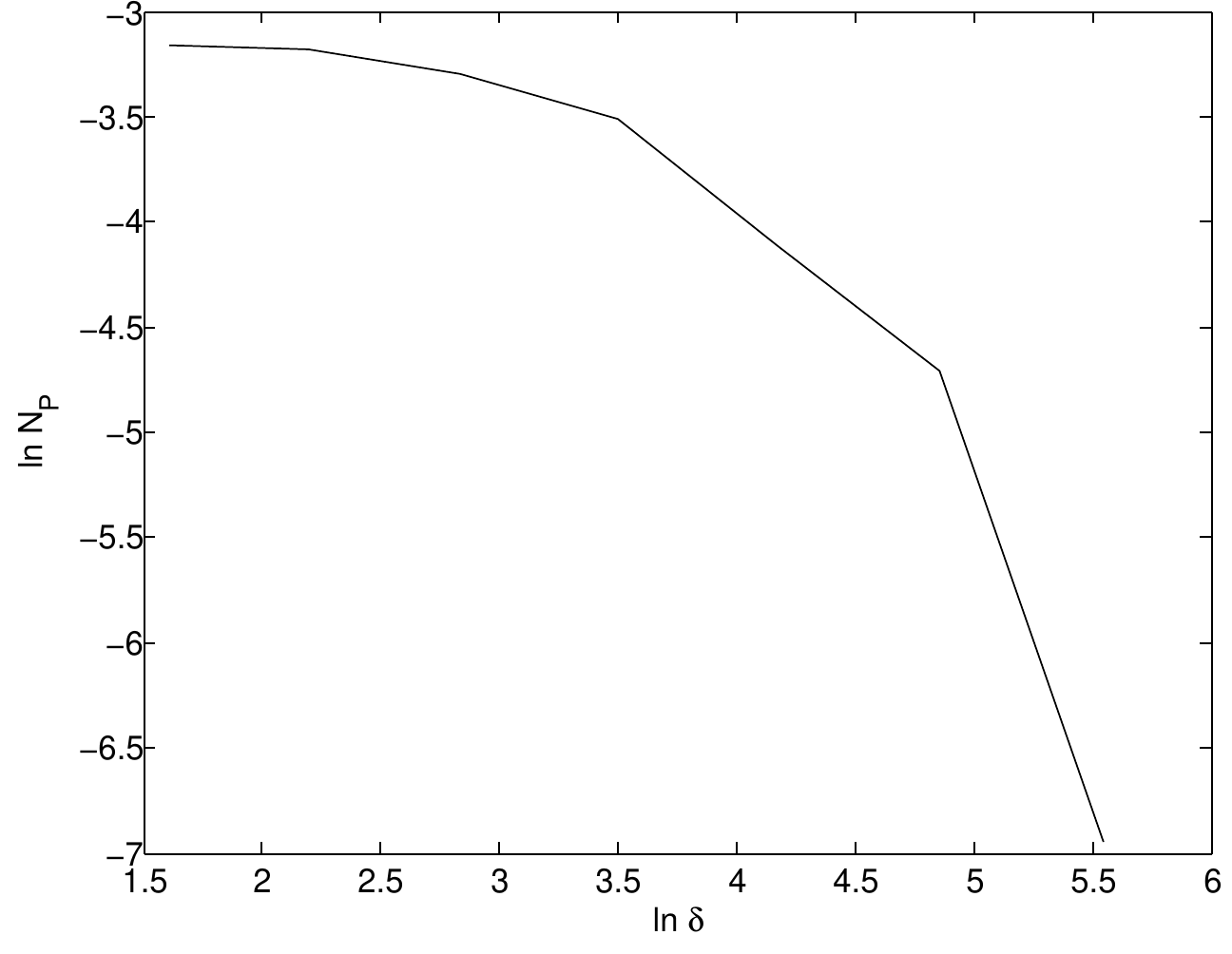}\\
		\includegraphics[width=0.3\textwidth]{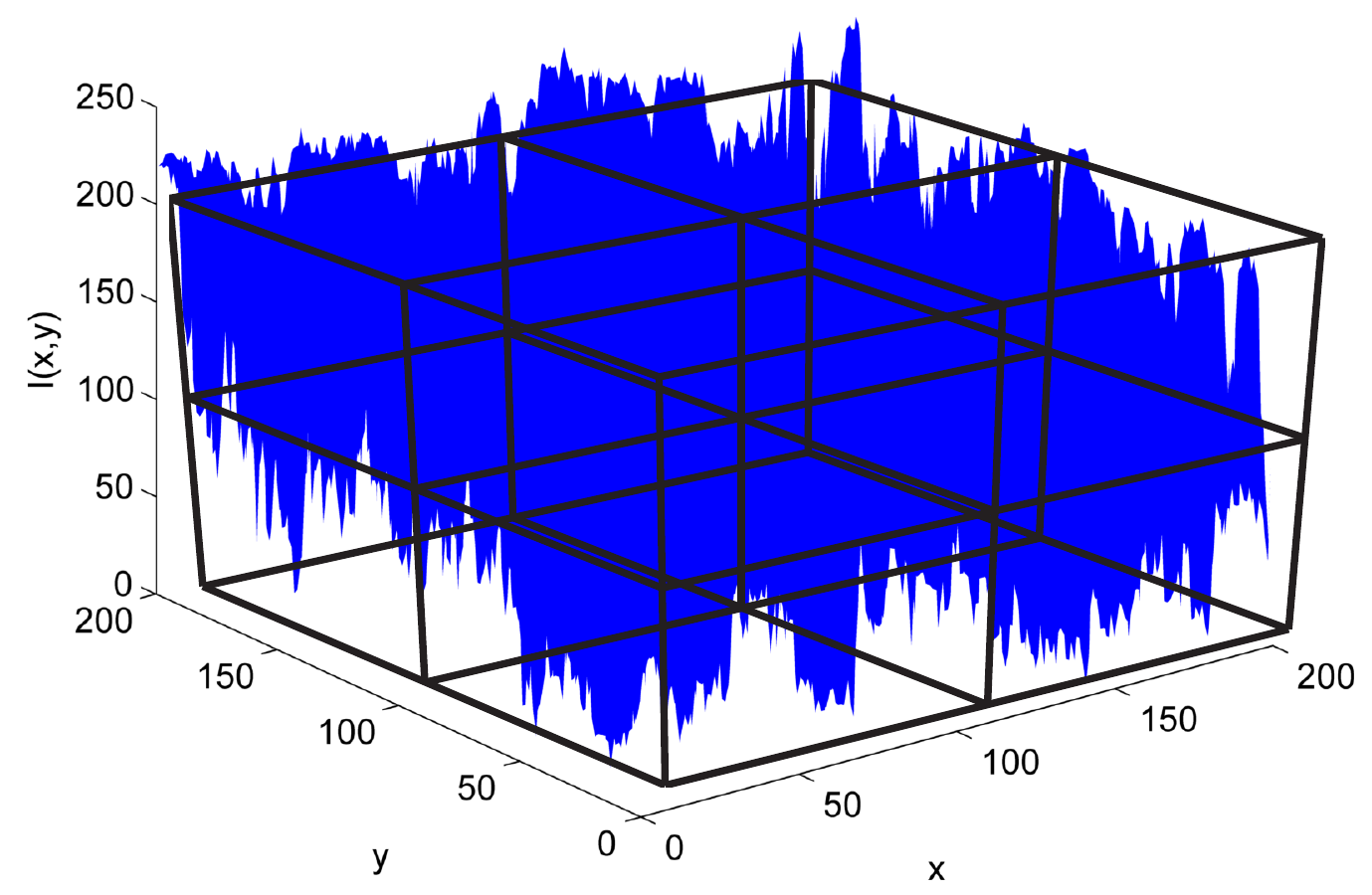} & \includegraphics[width=0.3\textwidth]{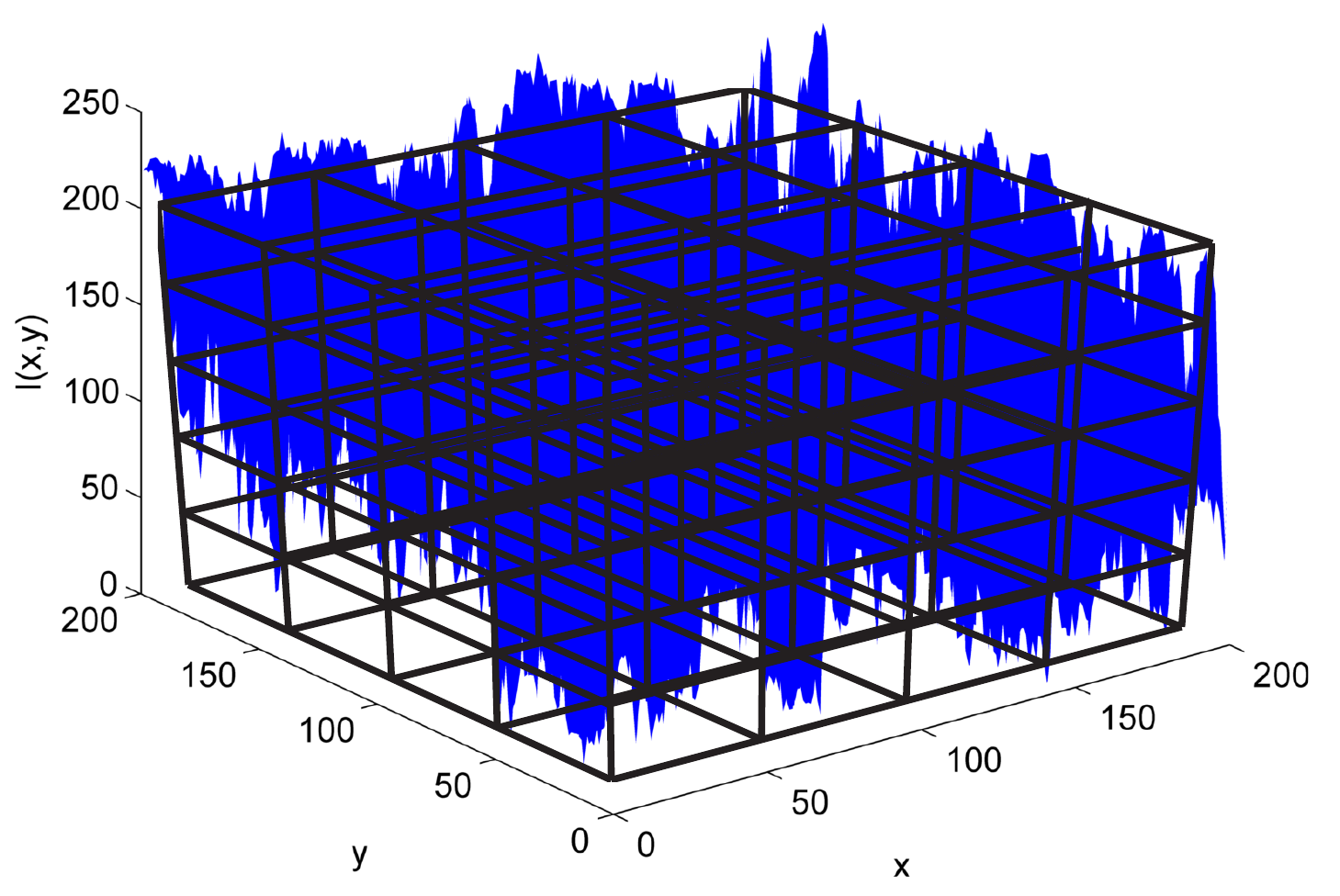} & \includegraphics[width=0.3\textwidth]{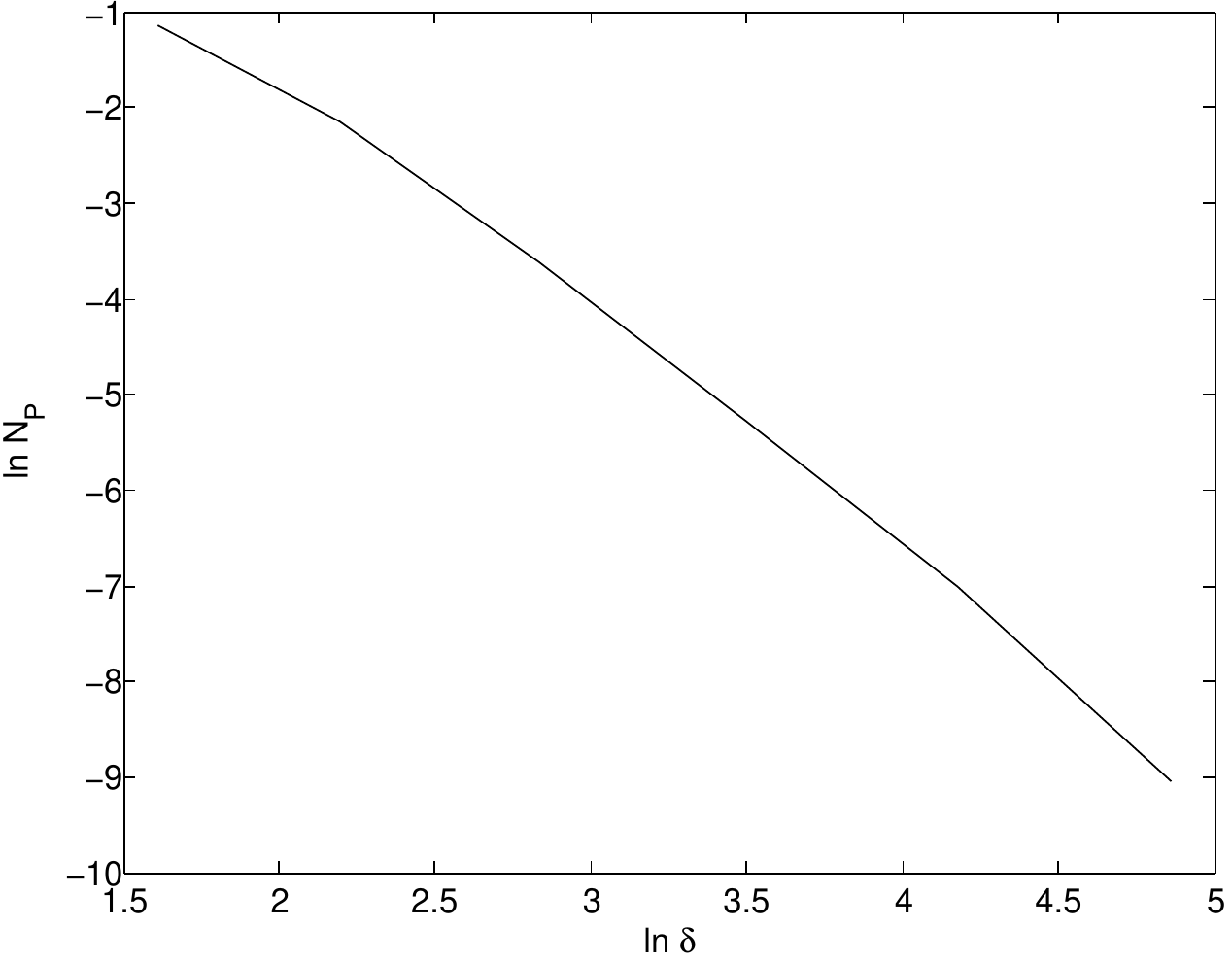}\\		
	\end{tabular}
	\caption{A diagram of probability dimension estimation. Above, the 2D version used for shapes analysis. Below, the 3D version used for gray-level images.}
	\label{fig:method}
\end{figure*}

\section{Fractal Descriptors}

The main idea of fractal descriptors is to extract values (descriptors) from the $\log-\log$ relation common to most methods of estimating fractal dimension. Actually, any fractal dimension method derived from Hausdorff dimension concept obey a power-law relation which may be explicit in the following:
\[\label{eq:FD2}
	D = -\frac{\log(\mathfrak{M})}{\log(\epsilon)},
\] 
where $\mathfrak{M}$ is a measure depending on the dimension method and $\epsilon$ is the scale under this measure is taken.

Therefore, fractal descriptors are provided from the the function $u$:
\[
	u:\log(\epsilon) \rightarrow \log(\mathfrak{M}).
\]
In order to simplify the notation we name the independent variable as $t$. Thus, $t = \log \epsilon$ and our fractal descriptors function is denoted $u(t)$. For the probability dimension used in this work, we have:
\[
	u(t) = -\frac{\log(N_V(\delta))}{\log(\delta)}.
\]

The values of function $u(t)$ may be used directly as descriptors of the analyzed image or may be post-processed by some kind of operation aiming at emphasizing some specifically aspects of that function. Here, we apply a multiscale transform to the function. In this way, we obtain a two-dimensional function $U(b,a)$, in which the variable $b$ is related to $t$ and $a$ is related to the scale on which the function is observed. A usual means of obtain $U$ is through a derivative process:
\[
	U(b,a) = \sum_{n = -\infty}^{+\infty}{u(t-n)G(n,a)},
\]
where $G$ is the well-known Gaussian function and $a$ is the smoothing parameter:
\[
	G(n,a) = \frac{1}{\sqrt{2\pi a}}\exp(-n^2/2t).
\]
Given the finite domain of function $u$, we also must restrict the response of Gaussian filter to a finite interval $[-M,+M]$:
\[
	U(b,a) = \sum_{n = -M}^{+M}{u(t-n)G(n,a)},
\]
where $M$ is a real value which should satisfy:
\[
	2\int_{t=M}^{\infty}G(t,a)dt = 2\int_{v=M/\sqrt{a}}^{\infty}G(v,a)dv < \mathfrak{e},
\]
being $\mathfrak{e}$ the tolerance error. Usually, we may have:
\[
	M = C\sigma+1 = C\sqrt(a)+1,
\]
where $C$ is a real constant with values commonly varying between $3$ and $6$.
\begin{figure}[!htpb]
\centering
\includegraphics[width=0.9\columnwidth]{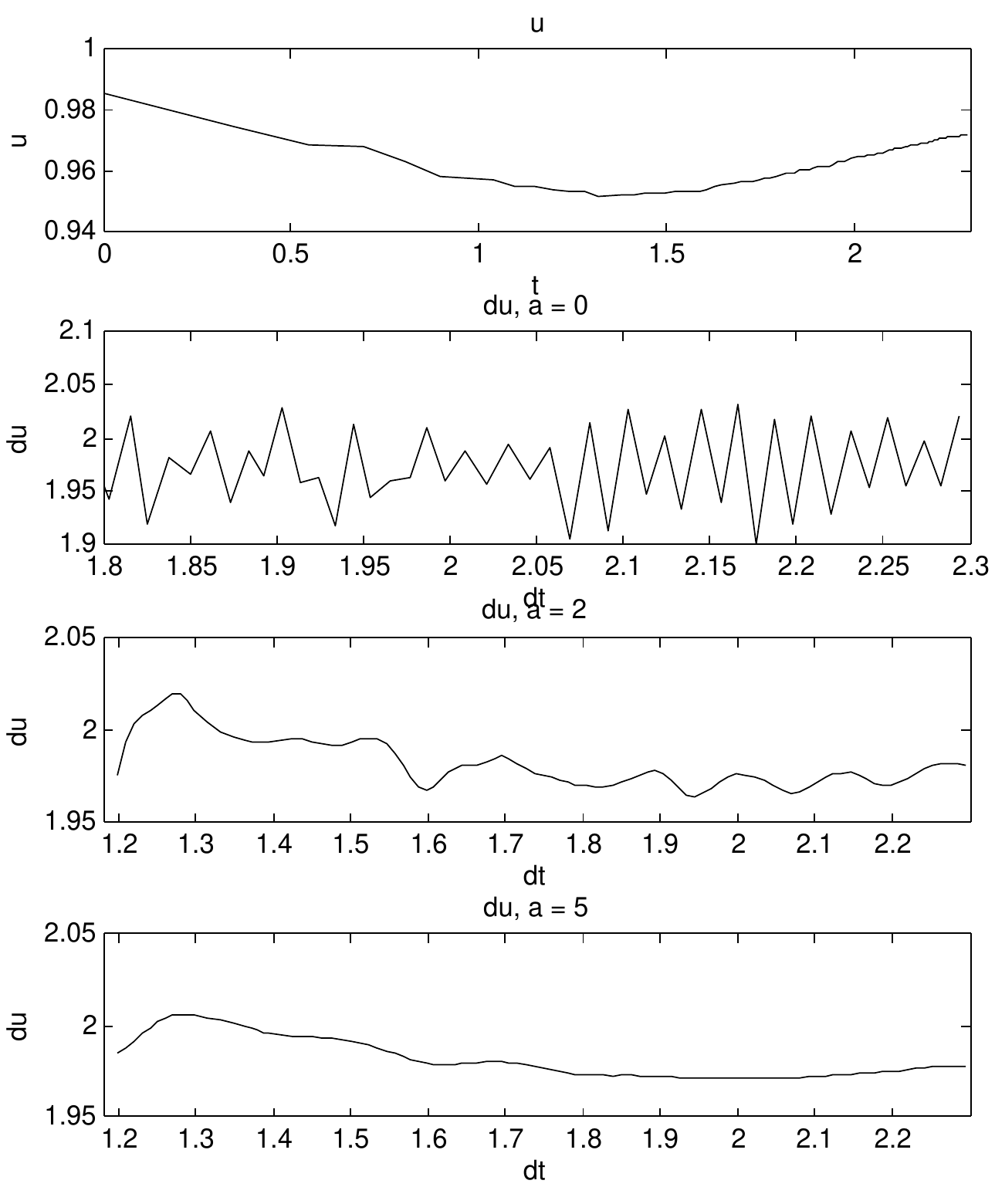}
\caption{Multiscale transform.}
\label{fig:multi}
\end{figure}

\section{Proposed Method}

The idea here proposed is to obtain fractal descriptors from textures based on the probability fractal dimension. Thus, such descriptors are computed from the curve $u(t):\log(N_P(\delta))$ in the Equation \ref{eq:prob}. Therefore, we apply a multiscale transform to $u$.

The multiscale process is achieved by deviating numerically and convoluting with a Gaussian filter, as described in the previous section:
\[
	U(b,a) = \frac{du}{dt} * G_a,
\]
where $G_a$ is the Gaussian function descritized over the interval $-10 \leq x \leq 10, x \in \mathbb{Z}$.

As multiscale transform maps a one-dimensional signal onto a two-dimensional function, it is a process which generates intrinsic redundancies. We may find different approaches to eliminate such redundancies keeping only the relevant information \cite{CC00}. Here, we adopt a simple method named fine-tuning smoothing in which $U(b,a)$ is projected under a specific value $a_0$ of the Gaussian parameter. Here we tested values of $a$ varying between $0.1$ and $5$ and used that values which provided the best performance in the experiments.

Finally, we selected a particular region from $U(b,a_0)$ to compose the descriptors. Empirically, we observed that the initial points in this curve are relevant to a good performance in our application. Then, we established a threshold $t$ after which all points in the convolution curve are disregarded. Thus, the values in the curve $U(i,a_0),1 \leq i \leq t$ are taken as the proposed descriptors.
\begin{figure}[!htpb]
\centering
\includegraphics[width=0.95\columnwidth]{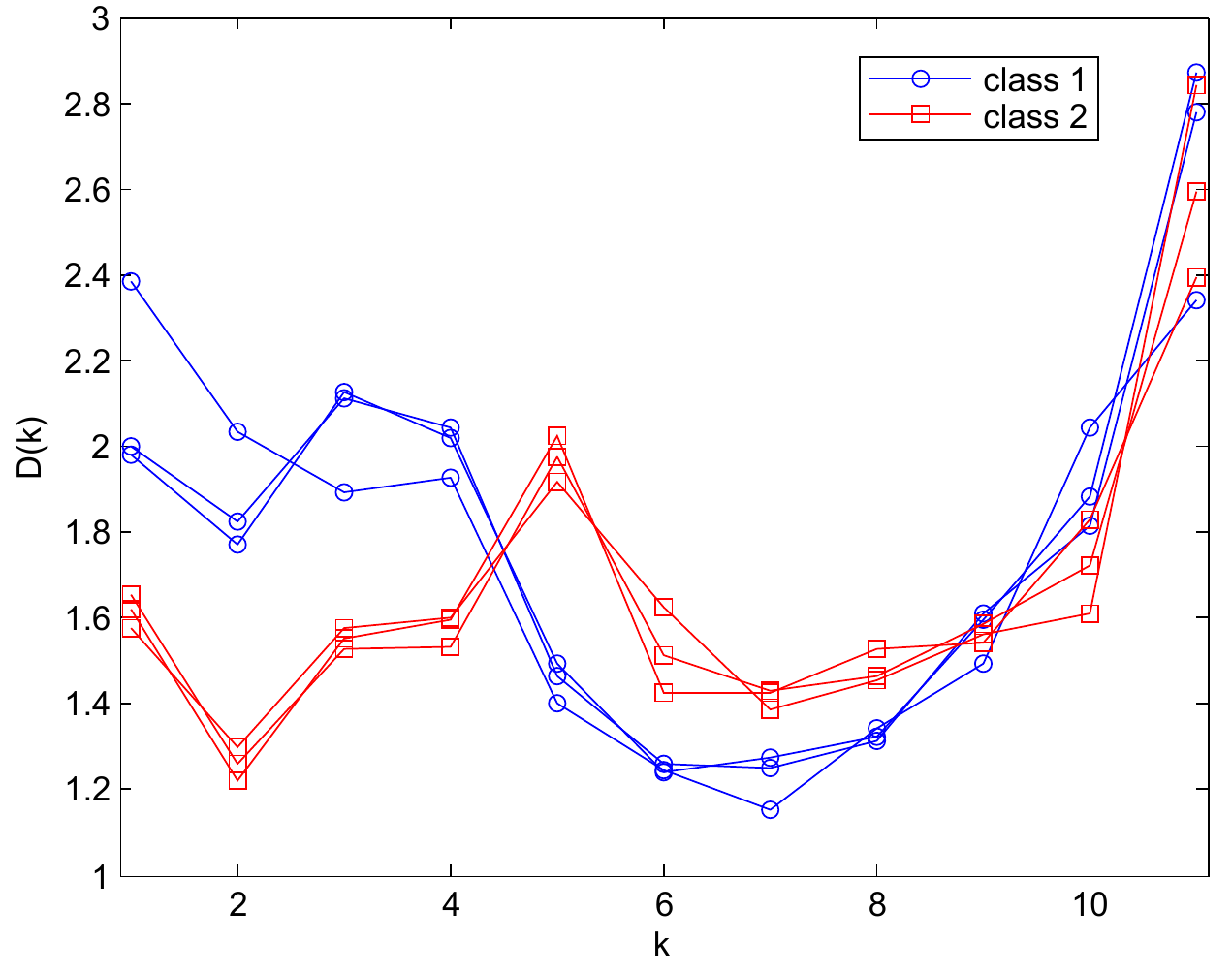}
\caption{Discrimination of texture image by the proposed descriptors.}
\label{fig:classes}
\end{figure}

\section{Experiments}

In order to verify the efficiency of the proposed technique, we applied the probability descriptors to the classification of two benchmark datasets and compared to the performance of other well-known and state-of-the-art methods for texture analysis.

The first classification was accomplished over the Brodatz dataset, a classical set of natural gray level textures photographed and put together in a book \cite{B66}. This dataset is composed by 111 classes with 10 textures with dimension 200$\times$200 in each class.

The second data set is the Vistex, a set of color textures extracted from natural scenes \cite{SS01}. Here, we employ a version of the dataset in which we have 7 classes, each one with a variable number of images of 256$\times$256 pixels and converted them to gray-level images.

We compare probability descriptors to other 4 other techniques, that is, Gabor wavelets \cite{MM96}, Co-occurrence matrix \cite{H67}, Gray Level Difference Method (GLDM) \cite{WDR76}, a multifractal approach described in \cite{PV02} and Bouligand-Minkowski fractal descriptors \cite{BCB09}.

Finally, we classify each descriptor by a hold-out process (half of data to train and the remaining to test) using K-Nearest Neighbor (KNN), with $K=1$ and compare the results.

\section{Results}

The Table \ref{tab:brod} shows the correctness rate in the classification of Brodatz dataset using the compared descriptors. The proposed method obtained the best result with a 12\% advantage. For this result we used $a = 0.1$ and threshold $t = 17$. A particular important aspect in Brodatz data set is the reduced number of descriptors of the proposed approach. This point may be specially important in large data basis when the computational performance is more relevant. Furthermore, the small number of features avoid the curse of dimensionality, which prejudices the reliability of the global result.
\begin{table}[!htpb]
	\centering
	\scriptsize
		\begin{tabular}{|c|c|c|}
			\hline
                 Method   & Correctness Rate (\%) & Number of descriptors\\
                 \hline
                  Gabor & 81.3 & 20\\
                  Co-occurrence & 53.5 & 84\\
                  GLDM & 52.2 & 20\\
                  Multifractal & 35.1 & 101\\
                  Bouligand-Minkowski & 47.6 & 85\\
                  Proposed method & 91.2 & 17\\
			\hline			
		\end{tabular}
	\caption{Correctness rate for Brodatz dataset.}
	\label{tab:brod}
\end{table}

On the other hand, the Table \ref{tab:vistex} shows the results for the Vistex textures. In this case, we obtained the best result by using $a = 0.1$ and $t = 80$. Again, the proposed approach provided the greater correctness with a 2\% advantage. Again, we have a good result in a data set which presents a lot of challenges once it is aimed at color analysis while the proposed approach is gray-level based. This aspect turns significant even a tiny classification enhancing.
\begin{table}[!htpb]
	\centering
	\scriptsize
		\begin{tabular}{|c|c|c|}
			\hline
                 Method   & Correctness Rate (\%) & Number of descriptors\\
                 \hline
                  Gabor & 85.7 & 20\\
                  Co-occurrence & 70.1 & 120\\
                  GLDM & 43.5 & 20\\
                  Multifractal & 20.1 & 101\\
                  Bouligand-Minkowski & 68.8 & 85\\
                  Proposed method & 87.7 & 80\\
			\hline			
		\end{tabular}
	\caption{Correctness rate for Vistex dataset.}
	\label{tab:vistex}
\end{table}

\begin{figure}[!ht] 
					 \centering
           \mbox{\subfigure[]{\epsfig{figure=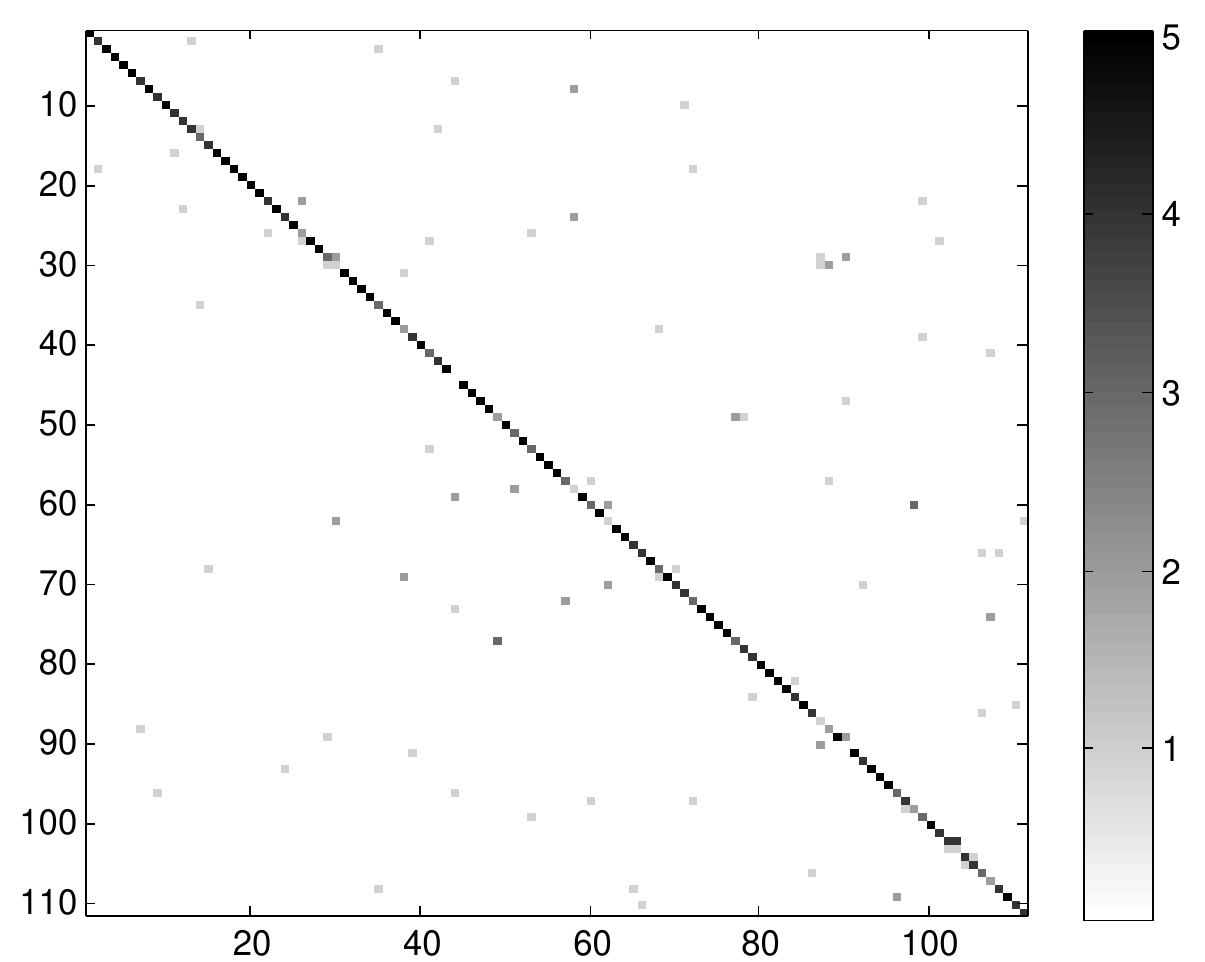,width=0.5\columnwidth}}
                 \subfigure[]{\epsfig{figure=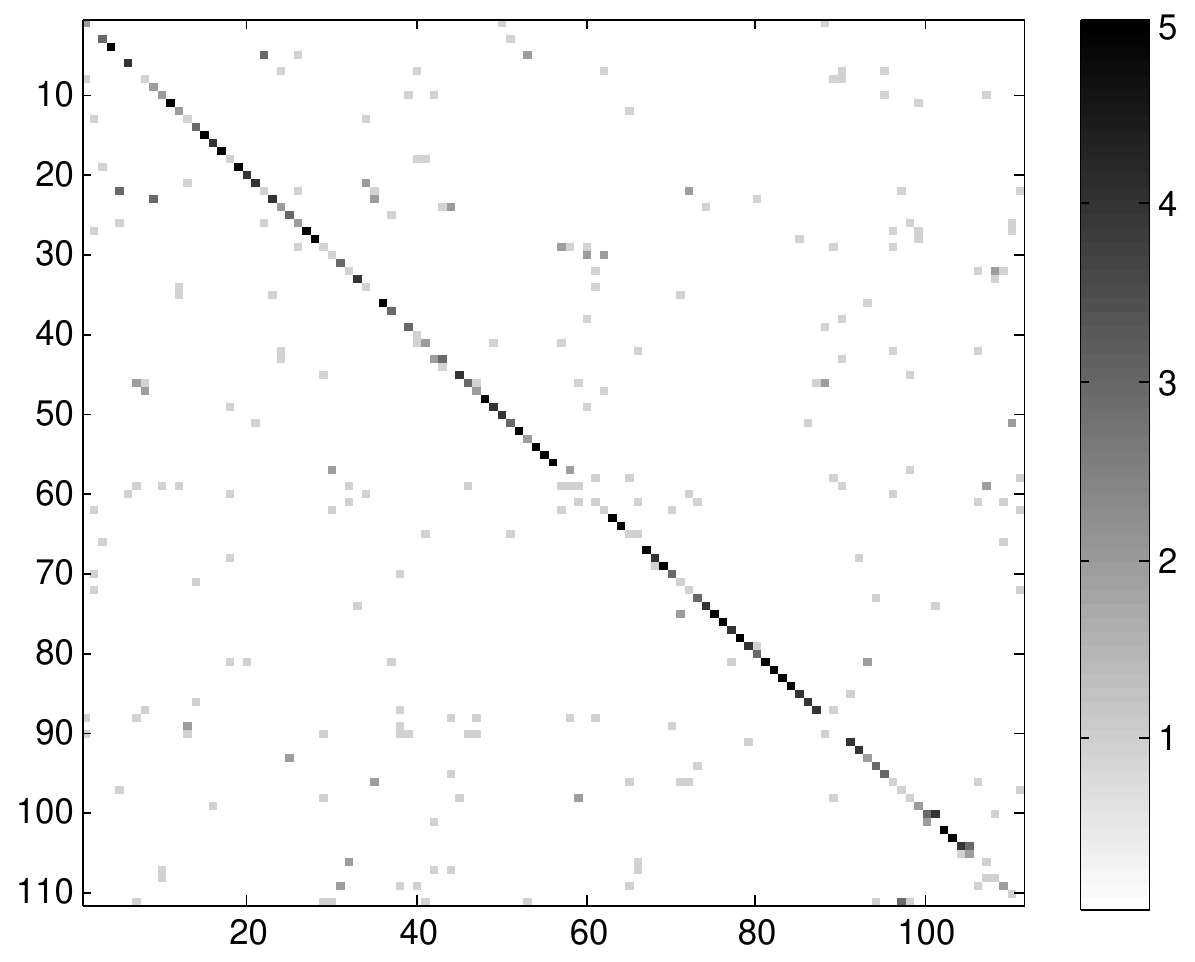,width=0.5\columnwidth}}}
           \mbox{\subfigure[]{\epsfig{figure=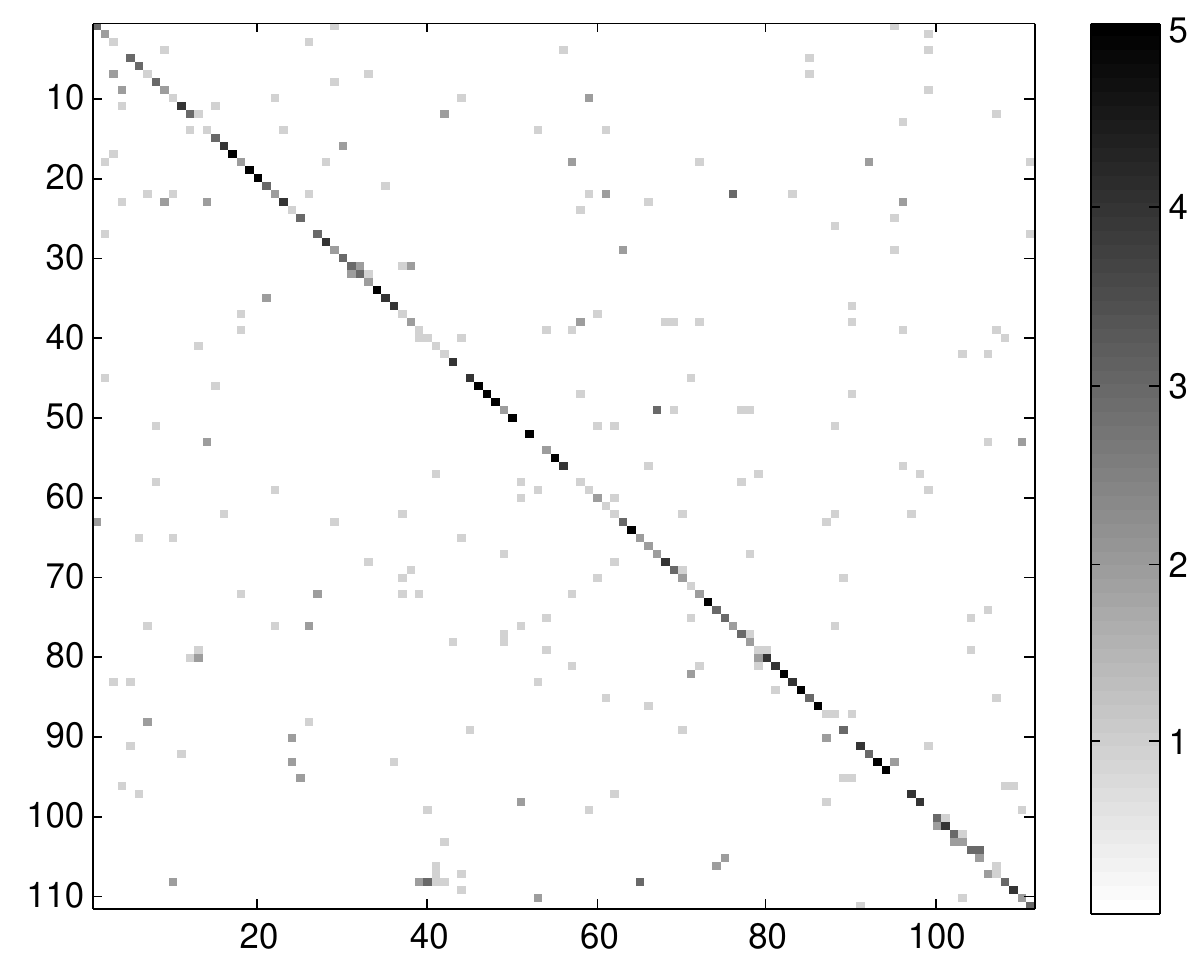,width=0.5\columnwidth}}
           			 \subfigure[]{\epsfig{figure=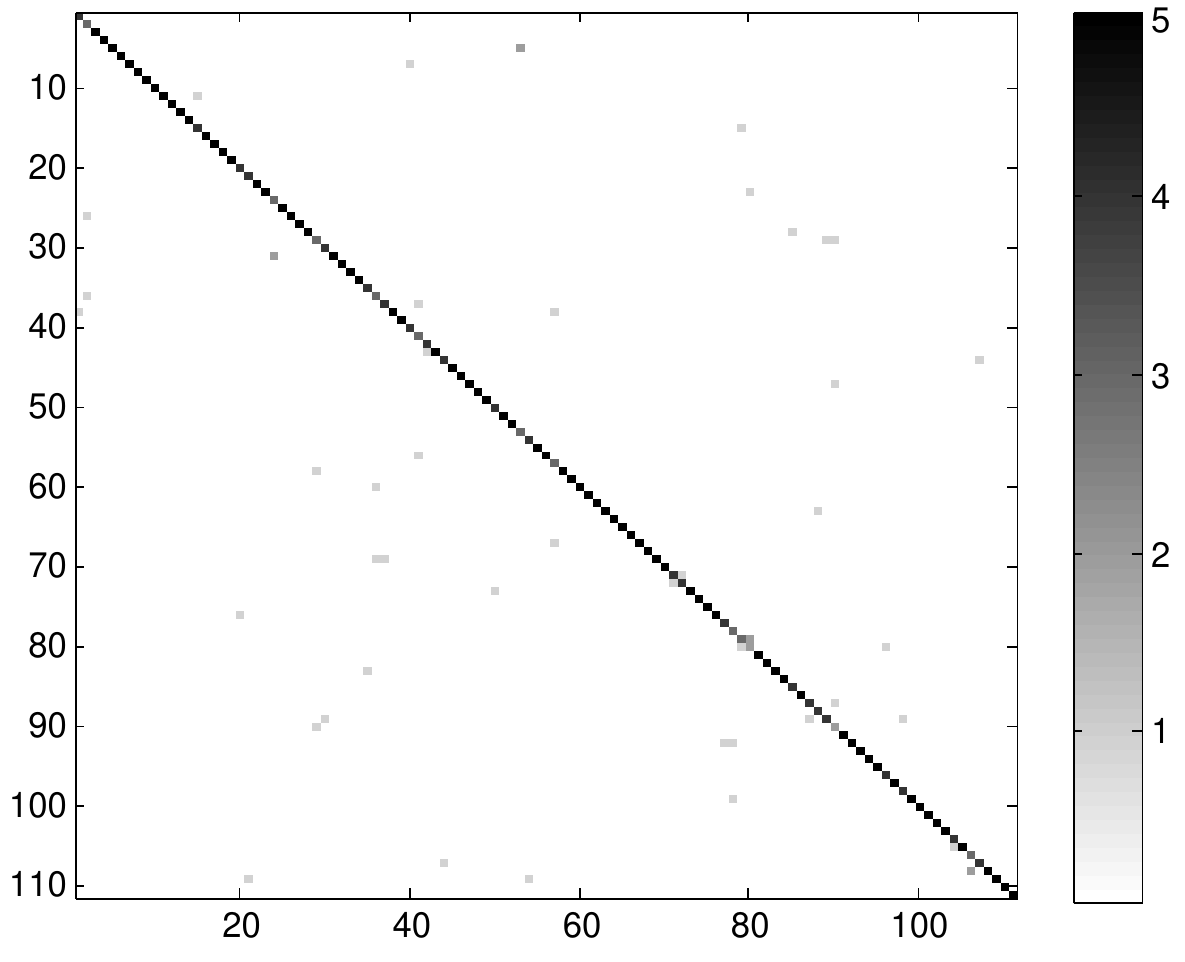,width=0.5\columnwidth}}}           			 
           \caption{Confusion matrices in Brodatz dataset. a) Gabor. b) Co-occurrence. c) GLDM. d) Proposed method. }
           \label{fig:CMbrodatz}                                  
   \end{figure}

\begin{figure}[!hb] 
					 \centering
           \mbox{\subfigure[]{\epsfig{figure=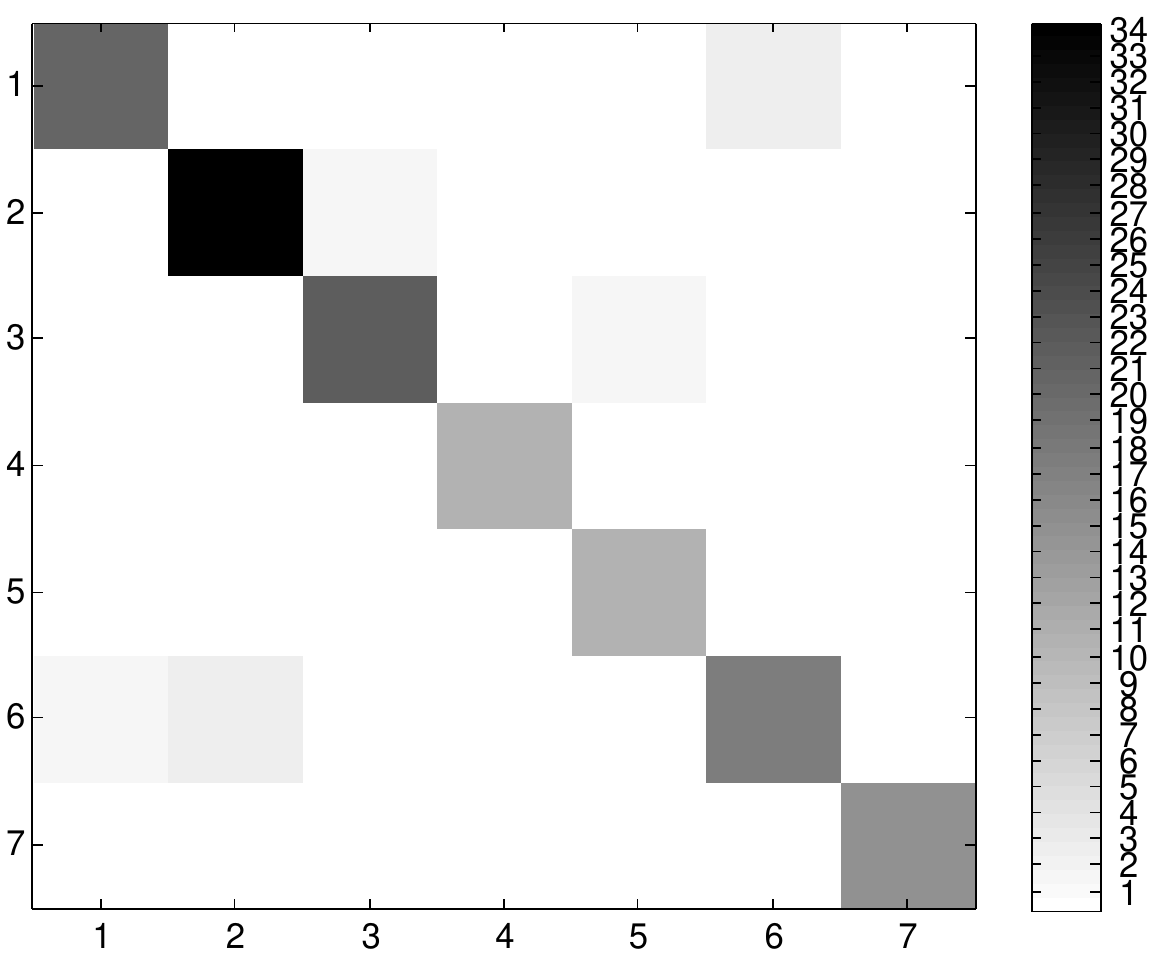,width=0.5\columnwidth}}
                 \subfigure[]{\epsfig{figure=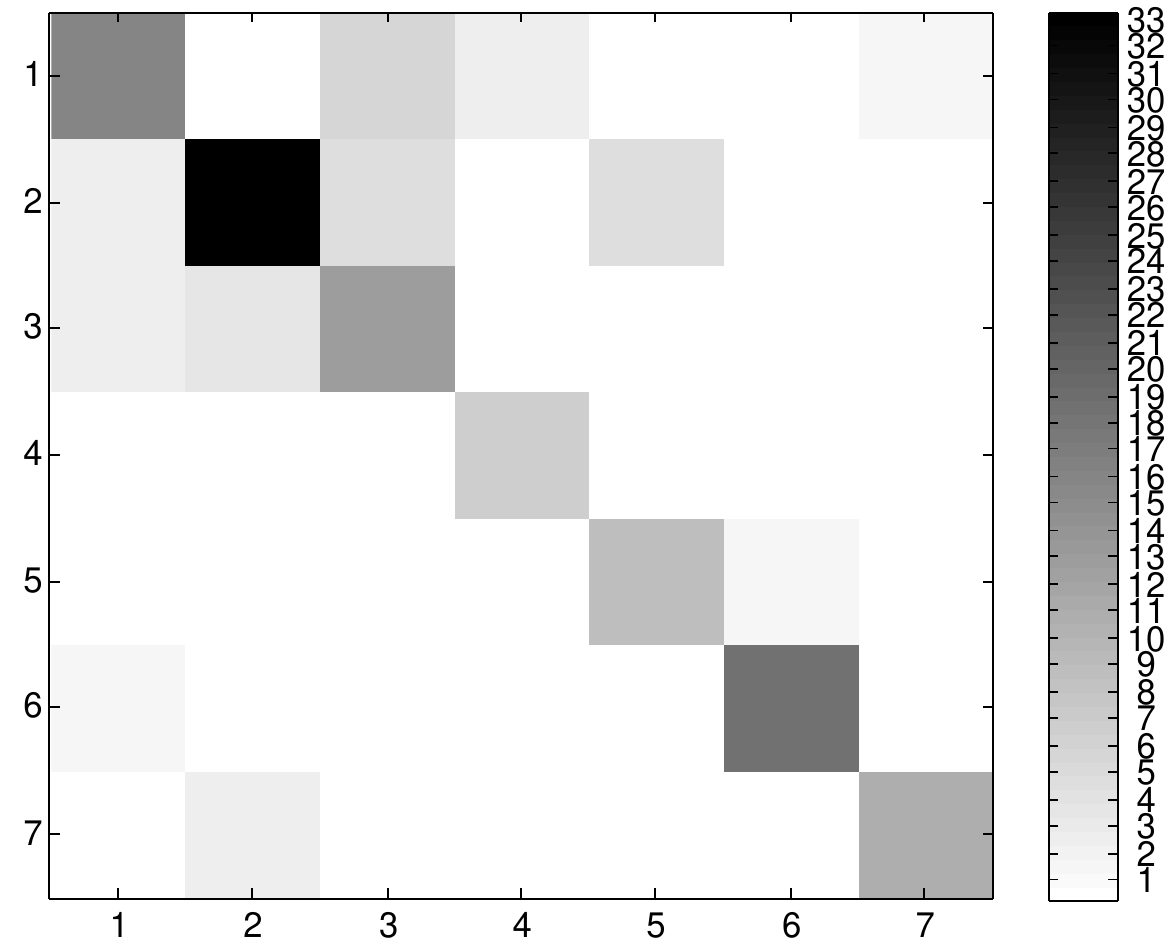,width=0.5\columnwidth}}}
           \mbox{\subfigure[]{\epsfig{figure=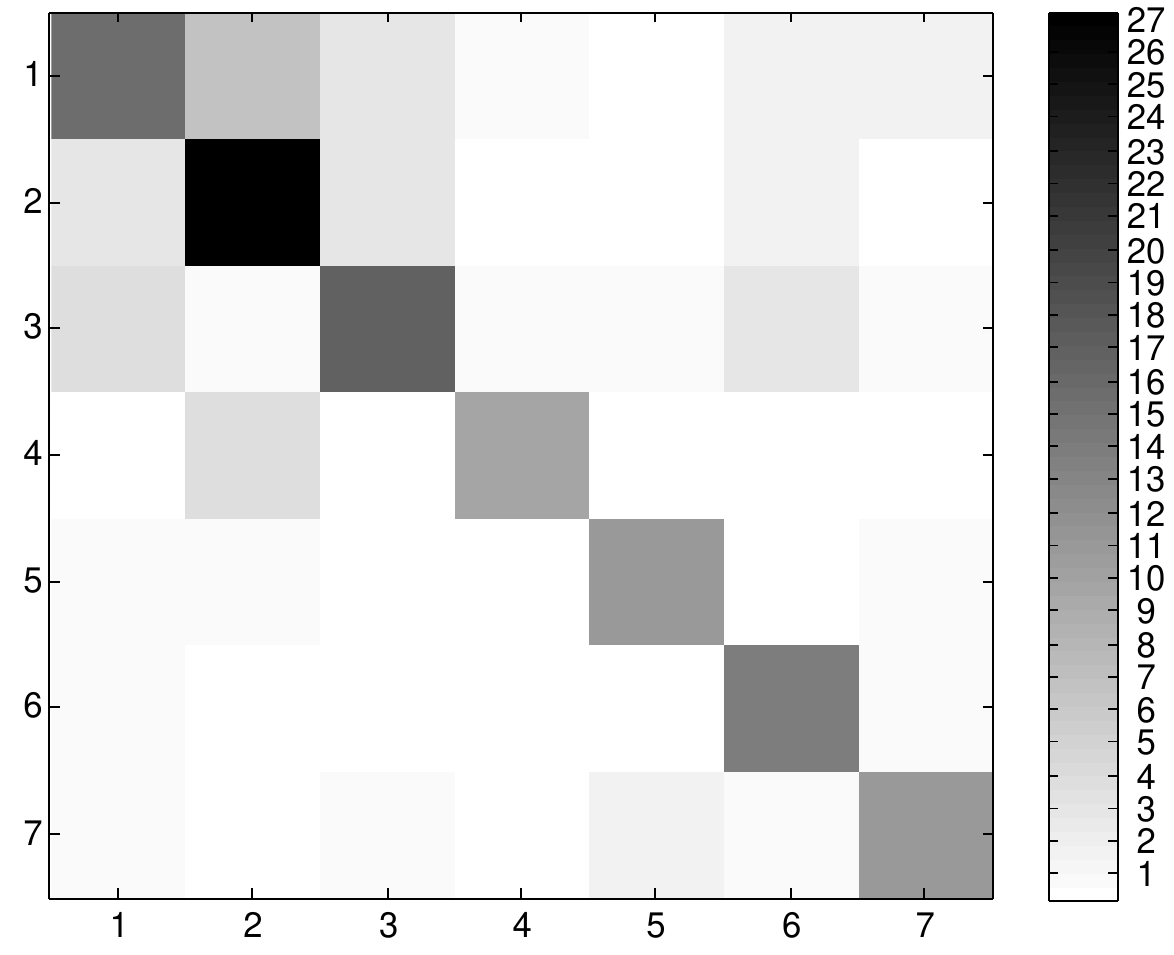,width=0.5\columnwidth}}
           			 \subfigure[]{\epsfig{figure=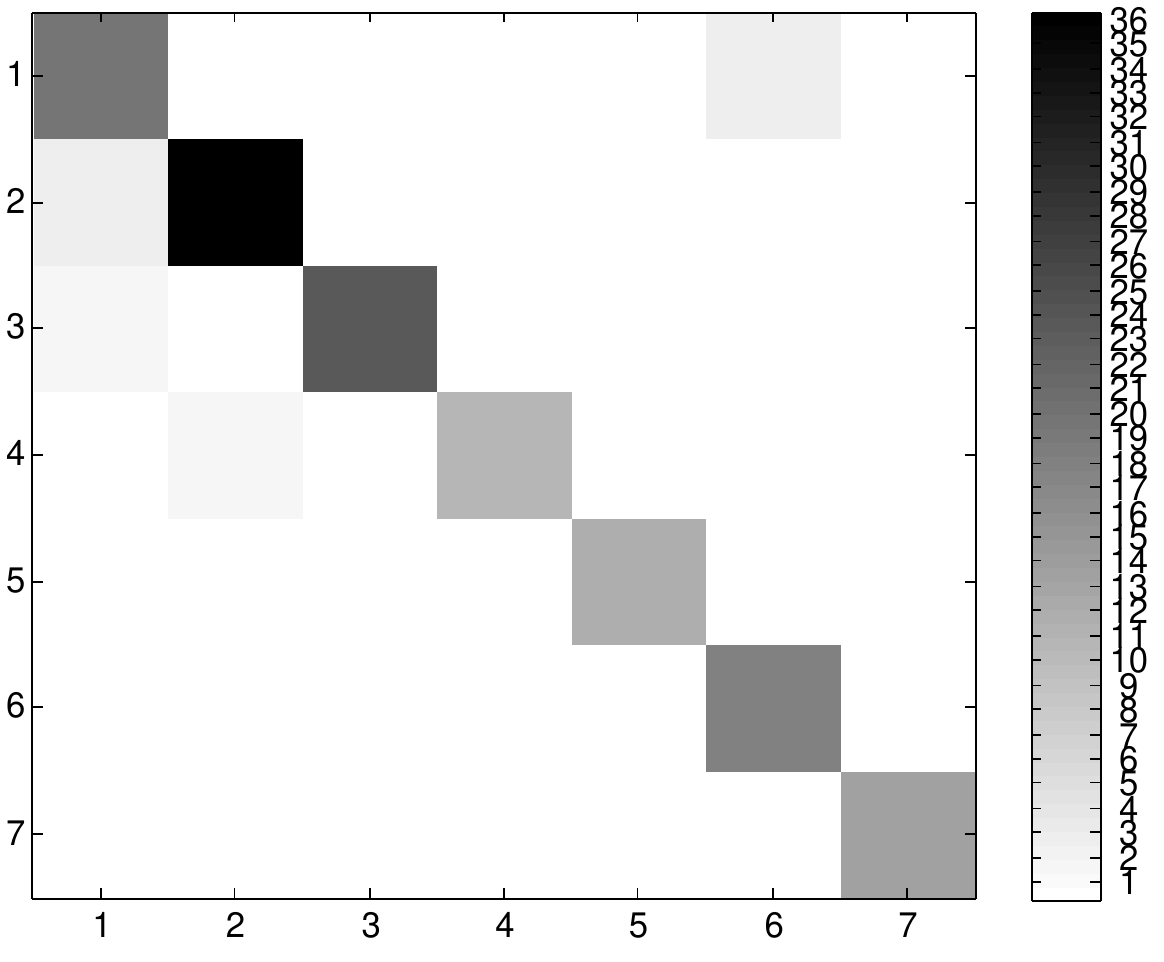,width=0.5\columnwidth}}}           			 
           \caption{Confusion matrices in Vistex dataset. a) Gabor. b) Co-occurrence. c) Bouligand-Minkowski. d) Proposed method. }
           \label{fig:CMvistex}                                  
   \end{figure}

Finally, the Figures \ref{fig:CMbrodatz} and \ref{fig:CMvistex} show the confusion matrices of the methods with best performances. In such figures, we have the predicted classes in the rows and the actual ones in the columns. The number of classes in each configuration is given by the intensity of gray-level in each point (brighter points correspond to large number of classes). In this kind of representation, a good descriptor must produce a matrix with a diagonal as brighter and continuous as possible and the minimum of brighter points outside the diagonal. In this sense, we see that, in Brodatz data, the probability descriptors presented exactly these characteristics, with almost no ``hole'' in the diagonal and with a lower density of brighter points outside.

In Vistex case, the matrices are not so distinguishable visually, even due to the similarity in the results. Thus, a perspective which we may use is analyze directly the number of corrected samples in each class. Based on this aspect, we observe in the legend bar that the proposed technique has its matrix normalized on a greater number of samples. This implies that although with a similar aspect, our approach presented a higher number of samples classified correctly in each class. 
   
A general analysis of the results demonstrates that the proposed method overcame the compared ones in both data sets, using a reasonable amount of descriptors. Such results was expected from fractal theory once we have a lot of works in the literature showing the efficiency of fractal geometry in the analysis of natural textures. Actually, fractal geometry presents a flexibility in the modeling of objects which cannot be well represented by Euclidean rules. The fractal dimension is a powerful metric for the complex patterns and spatial arrangement usually found in the nature. Fractal descriptors enhance such ability providing a way of capturing multiscale variations and nuances which could not be measured through conventional tools. More specifically, the probability descriptors here proposed aggregates a statistical approach to fractal analysis, composing a framework which supports a precise and reliable discrimination technique, as confirmed in the above results.     

\section{Conclusion}

The present work proposed a novel method to extract descriptors based on fractal theory for texture analysis application. 

Here we obtained such descriptors by applying a multiscale transform over the power law relation of fractal dimension estimated by the probability method.

We tested the efficiency of the novel technique in the classification of well-known benchmark texture dataset and compared its performance to that of other classical texture analysis methods. The results demonstrated that probability fractal descriptors are a powerful tool to model such textures. The method provides a rich way of representing even the most complex structures in texture images, being a reliable approach to solve a large class of problems involving the analysis of texture images.

\section*{Acknowledgments}

J.B.F. acknowledges support from CNPq.
O.M.B. acknowledges support from CNPq (Grant \#308449/2010-0 and \#473893/2010-0) and FAPESP (Grant \# 2011/01523-1). 


%

\end{document}